\documentclass[twocolumn]{article}

\usepackage[round]{natbib}
\usepackage{microtype}
\usepackage{graphicx}
\usepackage{subfigure}
\usepackage{booktabs}
\usepackage{hyperref}
\usepackage{xspace}
\usepackage{amsmath}
\usepackage{amssymb}
\usepackage{bm}
\usepackage{booktabs}
\usepackage{mathtools}
\usepackage{enumitem}
\usepackage{color}
\usepackage{physics}
\usepackage[T1]{fontenc}
\usepackage{hyperref}
\usepackage{wrapfig}
\usepackage{yfonts}
\usepackage{tikz}
\usepackage{balance}
\usepackage{mathptmx}
\usepackage{comment}
\usepackage{yfonts}
\usetikzlibrary{arrows, arrows.meta, positioning, shapes, fit, external}

\usepackage{cleanplots}
\include{preamble}

\newif\ifincludepdf
\includepdftrue

\begin{document}
	\frenchspacing
	
	\title{\textbf{Why did the distribution change?}} 
	
	\author{Kailash Budhathoki, Dominik Janzing, Patrick Bl\"obaum, and  Hoiyi Ng\\
		{\small Amazon}\\
		{\small \{kaibud, janzind, bloebp, nghoiyi\}@amazon.com} }
	
	\maketitle
	
	\begin{abstract}
		We describe a formal approach based on graphical causal models to identify the ``root causes'' of the change in the probability distribution of variables. After factorizing the joint distribution into conditional distributions of each variable, given its parents (the ``causal mechanisms''), we attribute the change to changes of these causal mechanisms.  
		This attribution analysis accounts for the fact that  mechanisms often change independently and sometimes only some of them change. 
		Through simulations, we study the performance of our distribution change attribution method. 
		We then present a real-world case study identifying the drivers of the difference in the income distribution between men and women.
	\end{abstract}

	\section{Introduction}\label{section:introduction}
	Changes to a probability distribution are common in many real-world domains that are part of a changing environment. For example, during COVID-19, most retailers most likely observed a shift in the distribution of their inventory level of products---
	as certain products have unusually high demand (e.g., masks and hand sanitizers) while supply for some other products are limited due to suppliers suspending manufacturing. To be able to effectively respond to (either proactively or retrospectively) similar situations, it is not only important to identify if and where the distribution changed, but also know \emph{why} the distribution changed. 
	
	In recent years, several techniques have been developed to either automatically detect changes in the underlying distribution from a sequence of observations~\citep{pollak:1985:dist-change-detection, kifer:2004:change-detection-data-stream}, or determine if two data samples come from the same distribution~\citep{chakravarti:1967:ks-test,scholz:1987:k-sample-test,snedecor:1989:two-sample-t-test,gretton:2012:kernel-two-sample-test}. However, a formal way to identify the ``root causes'' of the distribution change seems to be missing.
	
	In this work, we consider a system of $n$ variables $\rvX_1, \dotsc, \rvX_n$. In a supply chain, for instance, these variables can represent different business metrics, such as demand forecast, labour cost, shipment cost, and inventory level to name a few. 
	Typical \emph{causal} questions on these variables are interventional in nature, e.g. "What would be the impact on $\rvX_k$ if we were to intervene on $\rvX_j$?" Here we are interested in a slightly different question, namely "Which mechanisms are responsible for the change in the joint distribution $\dist_{\rvX_1, \dotsc, \rvX_n}$, or the marginal distribution $\dist_{\rvX_k}$ of one of the variables $\rvX_k$?" For example, in a supply chain, we might be interested in understanding the drivers of week-over-week changes in the distribution of inventory level (or a summary statistic, such as its mean across different products). To this end, we build upon graphical causal models~\citep{pearl:09:book}.

	Given a causal graph \causalGraph of variables $\rvX_1, \dotsc, \rvX_n$, assuming the causal Markov condition~\citep{spirtes:00:book}, we can factorise the joint distribution into causal conditionals, i.e.
	\begin{align*}
		\dist_{\rvX_1, \dotsc, \rvX_n} = \prod_{j=1}^{n} \dist_{\rvX_j \mid \PA_j},
	\end{align*}
	where $\dist_{\rvX_j \mid \PA_j}$ denotes the causal mechanism of variable $\rvX_j$ given its direct parents $\PA_j$ in the causal graph. Each parent-child relationship captured by $\dist_{\rvX_j \mid \PA_j}$ represents an autonomous physical mechanism---we can change one such relationship \emph{without} affecting the others.\!\footnote{This idea of autonomy of mechanisms  has a long history, see~\citet[Section 1.3]{pearl:09:book} and \citet[Section 2.2]{janzing:2017:causality-book} for historical notes.} 
	Thus it is plausible to attribute any change in the joint distribution or the marginal distribution of some target variable to the change in some of the causal mechanisms. Based on this insight, we develop a formal approach to define the quantitative contribution of each mechanism to the overall change. In particular, we use the Shapley value concept~\citep{shapley:1953:solution} from cooperative game theory to cope with the fact that the \emph{impact} of changing a mechanism depends on which other mechanisms have been changed already.
	
	The paper is structured as follows. In Section~\ref{section:mechanism-changes}, we formalise changes to causal mechanisms. Section~\ref{section:whyjoint} presents a proposal to attribute the change in the joint distribution. In Section~\ref{section:whymarg}, we describe a formal method to attribute the change in the marginal distribution to causal mechanisms. Section~\ref{section:detecting-mechanism-changes} discusses the practical implications of applying these attribution proposals.
	In Section~\ref{section:experiments}, we report results from simulations and present a case study in identifying the drivers of difference in the income distribution between men and women. Finally, we conclude in Section~\ref{section:conclusions}.
	
	\section{Causal model and mechanism changes}\label{section:mechanism-changes}
	We consider probabilistic causal models that incorporate probability to infer causal relationships between variables. Suppose that we have a collection of $n$ random variables $(\rvX_1, \dotsc, \rvX_n) \eqqcolon \mrvX$. The underlying causal graph \causalGraph of these variables is a directed acyclic graph in which a directed edge from $\rvX_i$ to $\rvX_j$ indicates that $\rvX_i$ causes $\rvX_j$ directly. A joint distribution $\dist_\mrvX$ is said to be compatible with the causal graph \causalGraph, if $\dist_\mrvX$ can be generated following the edges in \causalGraph. More formally, $\dist_\mrvX$ is compatible to \causalGraph if
	\begin{align}
		\dist_\mrvX = \prod_{j=1}^{n} \dist_{\rvX_j \mid \PA_j} \; .
	\end{align}
	
	\begin{definition}[Probabilistic Causal Model]
		A probabilistic causal model is a pair $\causalModel \coloneqq \langle \causalGraph, \dist_\mrvX \rangle$ that consists of a causal graph \causalGraph, and a joint distribution $\dist_\mrvX$ over the variables in \causalGraph that is compatible with \causalGraph.
	\end{definition}
	
	Given a probabilistic causal model, the causal Markov assumption~\citep{spirtes:00:book} allows us to factorise the joint distribution $\dist_\mrvX$ into causal mechanisms $\dist_{\rvX_j \mid \PA_j}$ at each node $\rvX_j$. Each causal mechanism $\dist_{\rvX_j \mid \PA_j}$ remains invariant to interventions (external influences) in other variables. With this, we can formally define what causal mechanism changes entail.
	
	\begin{definition}[Mechanism Changes]\label{def:mechanism-change}
		Mechanism changes to a causal model $\causalModel \coloneqq \langle \causalGraph, \dist_\mrvX \rangle$ on a subset of variables $\mrvX_T$ indexed by a change set $T \subseteq \{1, \dotsc, n\}$ transform \causalModel into $\causalModel_T \coloneqq \langle \causalGraph, \dist^{T}_\mrvX \rangle$, where 
		\[
		\dist^{T}_\mrvX = \prod_{j \in T} \newdist_{\rvX_j \mid \PA_j} \prod_{j \notin T} \dist_{\rvX_j \mid \PA_j} \; 
		\]
		is a new joint distribution obtained by replacing ``old'' causal mechanism $\dist_{\rvX_j \mid \PA_j}$ at each node $\rvX_j$, where $j \in T$, with the ``new'' causal mechanism $\newdist_{\rvX_j \mid \PA_j}$.
	\end{definition}
	
	The following example illustrates the idea above in a formal setting where causal relationships between variables are represented in terms of structural equations~\citep{pearl:09:book}.
	
	\begin{example}
		Consider a causal model consisting of two variables, i.e. $\causalModel = \langle \rvX_1 \rightarrow \rvX_2, \dist_{\rvX_1, \rvX_2} \rangle$, induced by the structural equations $\rvX_1 \coloneqq \rvN_1$ and $\rvX_2 \coloneqq 2 \rvX_1 + \rvN_2$, where the independent unobserved noise terms $N_j \sim \mathcal{N}(0, 1)$ are distributed according to a standard Normal distribution. A typical structure preserving intervention~\citep{eberhardt:2007:interventions} changes either the parent-child functional 
		relationship, or the distribution of unobserved noise term. Consider an intervention that changes the relationship between $\rvX_1$ and $\rvX_2$ from the linear function to a non-linear function represented by the new structural assignment $\rvX_2 \coloneqq \rvX_1^3 + \rvN_2$. Whereas this changes the causal mechanism of $\rvX_2$ from $\dist_{\rvX_2 \mid \rvX_1}$ to $\newdist_{\rvX_2 \mid \rvX_1}$, the underlying causal graph remains the same. As such, we have a new causal model $\causalModel_{\{2\}} = \langle \rvX_1 \rightarrow \rvX_2, \newdist_{\rvX_1, \rvX_2} \rangle$, where $\dist^{\{2\}}_{\rvX_1, \rvX_2} = \dist_{\rvX_1} \newdist_{\rvX_2 \mid \rvX_1}$.
	\end{example}
	
	The ``new'' joint distribution $\dist^{T}_\mrvX$ can also be seen as the post-intervention joint distribution $\dist^{\mathit{do}(T)}_\mrvX$ where $\mathit{do}(T)$ represents mechanism changes through stochastic intervention at each node $\rvX_j$ indexed by $T$~\citep{correa:2020:stochastic-interventional-calculus}. In particular, we only consider the cases where the change in the joint distribution is a result of systematic replacements of independent physical mechanisms. Other cases, e.g. adversarial perturbation, can also change the joint distribution arbitrarily, which we do not consider here. Moreover, we consider the problem setting where the joint distribution changes, but \emph{not} the causal graph.
	
	\section{Why did the joint distribution change?}\label{section:whyjoint}
	As the joint distribution $\dist_{\rvX_1, \dotsc, \rvX_n}$ is a composition of independent causal mechanisms $\dist_{\rvX_j \mid \PA_j}$, it is plausible to attribute any change in the joint distribution to the change in some of the causal mechanisms. We would like to compute the contribution of each node---potentially due to the change in its causal mechanism---to the change in the joint distribution. 
	
	To this end, first we need a measure that quantifies the change in the joint distribution. A natural choice for quantifying the change in the joint distribution are the divergence measures as they measure the ``distance'' between two probability distributions. 
	In this work, we consider the Kullback-Leibler (KL) divergence (also called relative entropy)~\citep{cover:2006:information-theory}. 
	Let \dist and $Q$ be two probability distributions of a discrete random variable defined on the same probability space \domainX. Then the KL divergence from $Q$ to \dist is defined as
	\begin{align*}
		\kld{\dist}{Q} \coloneqq \sum_{\obsRvX \in \domainX} \dist(\obsRvX) \log \left( \frac{\dist(\obsRvX)}{Q(\obsRvX)} \right).
	\end{align*}
	If \dist and $Q$ are the distributions of a continuous random variable, then the summation is replaced by an integral. The KL divergence is particularly suitable for our purpose as it is additive for the independent compositions of the joint distribution. That is, by generalising the chain rule to more than two variables using the causal Markov condition, we get an additive decomposition of the KL divergence from the joint distribution $\dist_\mrvX$ to $\newdist_\mrvX$~\citep[Theorem 2.5.3]{cover:2006:information-theory} :
	\begin{align*}
		\kld{\newdist_\mrvX}{\dist_\mrvX} = \sum_{j=1}^{n} \kld{\newdist_{\rvX_j \mid \PA_j}}{\dist_{\rvX_j \mid \PA_j}}
	\end{align*}
	Thus, the contribution of each node $\rvX_j$ to the KL divergence from the joint distribution $\dist_\mrvX$ to $\newdist_\mrvX$ is the KL divergence from its causal mechanism $\dist_{\rvX_j \mid \PA_j}$ to $\newdist_{\rvX_j \mid \PA_j}$. In other words, each node---due to the change in its causal mechanism---contributes independently to the KL divergence from the joint distribution $\dist_\mrvX$ to $\newdist_\mrvX$. The lemma below formalises this observation.
	\begin{definition}
		Suppose that the causal mechanism of a node $\rvX_j$ changes from $\dist_{\rvX_j \mid \PA_j}$ to $\newdist_{\rvX_j \mid \PA_j}$. Then the contribution of a node $\rvX_j$ to the KL divergence from the joint distribution $\dist_\mrvX$ to $\newdist_\mrvX$ is $\kld{\newdist_{\rvX_j \mid \PA_j}}{\dist_{\rvX_j \mid \PA_j}}$.
	\end{definition}
	The KL divergence is always non-negative. That is, for any \dist and $Q$, it holds that $\kld{\dist}{Q} \geq 0$. Therefore, the contribution of a node $\rvX_j$ to the change in the joint distribution, measured in terms of the KL divergence, \emph{cannot} be negative. If the causal mechanism of a variable did \emph{not} change, then its contribution will be zero. 
	
	We should add a remark, however, that the KL divergence between conditionals also depends on the distribution of parents, not only the conditionals. Formally, the KL divergence from $\dist_{\rvX_j \mid \PA_j}$ to $\newdist_{\rvX_j \mid \PA_j}$ is defined as
	\[
	\kld{\newdist_{\rvX_j \mid \PA_j}}{\dist_{\rvX_j \mid \PA_j}} \coloneqq \exptdist{\PA_j \sim \newdist_{\PA_j} }{\kld{\newdist_{\rvX_j \mid \pa_j}}{\dist_{\rvX_j \mid \pa_j}}}.
	\] 
	This is not really problematic, however, as the marginal distribution  $\newdist_{\PA_j}$ of the parents is only used for averaging the KL divergences $\kld{\newdist_{\rvX_j \mid \pa_j}}{\dist_{\rvX_j \mid \pa_j}}$. As such, each node $\rvX_j$ uses the corresponding parent-distribution $\newdist_{\PA_j}$ from the same joint distribution $\newdist_{\mrvX}$. 
	
	Estimating KL divergence in high-dimensional setting is a challenging problem. For some parametric families, such as the exponential family of distributions,\!\footnote{The exponential family of distributions includes the Gaussian, Poison, Binomial, Multinomial, and Beta, as well as many others.} however, closed-form expressions exist for computing KL divergence. If a non-parametric estimator is desired, then we can use the $k$-nearest-neighbour-based estimator of KL divergence~\citep{wang:2009:k-nearest-kl-divergence-estimator} that is asymptotically unbiased and mean-square consistent assuming i.i.d. samples.
	
	One may also wonder whether the asymmetry of KL divergence creates a non-intuitive interpretation---as the impact of each mechanism change differs according to the direction of comparison. From an inferential standpoint, however, one direction seems preferable. When a distribution changes in time, there is a natural inferential asymmetry since one would rather consider the likelihood of new data with respect to the old model than vice versa. However, the main reason to choose KL divergence is that it nicely decomposes additively. Admittedly, this additive decomposition is a bit spoiled because we need a reference distribution to weigh the change of the conditionals---but this seems like a problem that is hard if not impossible to avoid.
	
	In practice, often it is of interest to understand why the \emph{marginal} distribution of one \emph{target} variable changed, instead of the change in the joint distribution of all variables. In the next section, using a concept from game theory, we formalise how to attribute the change in the marginal distribution of a target variable to each node in the causal graph.
	
	\section{Why did the marginal distribution change?}\label{section:whymarg}
	Suppose that the marginal distribution of a target variable $\rvX_k$ changes---from $\dist_{\rvX_k}$ to $\newdist_{\rvX_k}$.
	The causal Markov condition allows us to compute the marginal distribution of any variable in the causal graph by marginalising (summing) over all independent causal mechanisms excluding that of the variable itself. Formally, given a causal model $\causalModel = \langle \causalGraph, \dist_{\mrvX} \rangle$, the marginal distribution $\dist_{\rvX_k}$ of the variable $\rvX_k$ can be computed by first factorising the joint distribution using the causal Markov condition, and then marginalising over all other variables, i.e.
	\begin{align*}
		\dist_{\rvX_k} &= \sum_{x_1, \dotsc, x_{k-1}, x_{k+1}, \dotsc, x_n} \dist_{\rvX_1, \dotsc, \rvX_n} \\
		&= \sum_{x_1, \dotsc, x_{k-1}, x_{k+1}, \dotsc, x_n} \prod_{j=1}^{n} \dist_{\rvX_j \mid \PA_j}
	\end{align*}
	The change in the marginal distribution of $\rvX_k$ given the change set $T$ is then given by
	\begin{align*}
		\dist^T_{\rvX_k} = &= \sum_{x_1, \dotsc, x_{k-1}, x_{k+1}, \dotsc, x_n} \prod_{j \in T} \newdist_{\rvX_j \mid \PA_j} \prod_{j \notin T} \dist_{\rvX_j \mid \PA_j} .
	\end{align*}
	
	It is therefore reasonable to assume that any change in the marginal distribution $\dist_{\rvX_k}$ of the variable $\rvX_k$ is most likely due to the change in some of the causal mechanisms $\dist_{\rvX_j \mid \PA_j}$. Unlike in case of the joint distribution change, however, the additive property of KL divergence cannot be leveraged directly for the attribution here---due to the marginalisation. 
	
	A natural way to compute the contribution of each node to the change in the marginal distribution of the target, from $\dist_{\rvX_k}$ to $\newdist_{\rvX_k}$, is then as follows: replace each ``old'' mechanism $\dist_{\rvX_j \mid \PA_j}$ by the ``new'' mechanism $\newdist_{\rvX_j \mid \PA_j}$ in succession. Each replacement changes the marginal distribution of the target $\rvX_k$, which can be used to compute the contribution of the corresponding node. The amount of change, however, depends on the causal mechanisms that have been already replaced. In other words, the contribution of a node $\rvX_j$ to the change in the marginal distribution of the target $\rvX_k$ depends on the order in which we replace the causal mechanisms---the resulting attribution procedure is hence in danger of becoming arbitrary.
	
	\subsection{Shapley Values}
	The Shapley value~\citep{shapley:1953:solution} from cooperative game theory provides a principled approach to mitigate the arbitrariness in the attribution procedure, due to the dependence on the ordering of replacements. In particular, it removes the arbitrariness by symmetrizing over all orderings. We briefly summarise the Shapley value here.
	
	Let $N \coloneqq \{1, \dotsc, n\}$ be a set of $n$ players and $\setFunction:2^N \rightarrow \mathbb{R}$ be a set function that associates a real-valued payoff to a coalition of players $\coalition \subseteq N$ with $\setFunction(\emptyset)=0$, where $\emptyset$ denotes an empty set. We assume that players will cooperate to form a grand coalition $N$. The goal is then to ``fairly'' assign the resulting payoff $\setFunction(N)$ to each player $j$ in $N$.
	
	Let $\permutation: N \rightarrow N$ denote a permutation of players $N$. All permutations of the set $N$ with $n$ elements form a symmetric group $S_n$. 
	Suppose that each player enters into a coalition one by one in the ordering $\permutation(1), \permutation(2), \dotsc, \permutation(n)$ to eventually form a grand coalition. Let $\precedingPlayers(j)$ denote the set of players that precede player $j$ in the ordering, i.e. $\precedingPlayers(j) \coloneqq \{ i \in N \mid \permutation^{-1}(i) < \permutation^{-1}(j) \}$. Note that $\sigma^{-1}(i)$ gives player $i$'s position. Then the change in the payoff of a coalition when a player joins the coalition is the marginal contribution of the player to the coalition,
	\begin{align*}
		\marginalContribution{\setFunction}{j \mid \precedingPlayers(j)} \coloneqq \setFunction\big(\precedingPlayers(j) \cup \{j\}\big) - \setFunction\big(\precedingPlayers(j)\big) \; .
	\end{align*}
	The marginal contributions of all players will then sum up to the payoff of the grand coalition $\setFunction(N)$, i.e.
	\begin{align}
		\sum_{j = 1}^{n} \marginalContribution{\setFunction}{j \mid \precedingPlayers(j)} = \setFunction(N) \; .
	\end{align}
	Unfortunately, the marginal contribution of a player $j$ depends on the order \permutation in which players enter into a coalition. To remove the dependence on the ordering \permutation, Shapley's solution~\citep{shapley:1953:solution} is to assign each player $j \in N$ its average marginal contribution over all permutations, i.e.
	\begin{align*}
		\shapleyValue_j(\setFunction)
		&\coloneqq \frac{1}{n!} \sum_{\sigma \in S_n} \setFunction\big(\precedingPlayers(j) \cup \{j\}\big) - \setFunction\big(\precedingPlayers(j)\big)\\
		&= \sum_{\coalition \subset N\setminus 
			\{j\}} \frac{|\coalition|! (n-|\coalition|-1)!}{n!} \left( \setFunction(\coalition \cup \{j\}) - \setFunction(\coalition) \right)\\
		&= \sum_{\coalition \subset N\setminus 
			\{j\}} \frac{1}{n \binom{n-1}{|T|}} \left( \setFunction(\coalition \cup \{j\}) - \setFunction(\coalition) \right) \; ,
	\end{align*}
	where the second line follows from counting the number of permutations (out of $n!$) where $\precedingPlayers(j)=\coalition$. The Shapley value is also desirable because it gives a unique solution to the following axioms that capture the notion of fairness.
	
	\begin{description}[leftmargin=0pt]
		\item[Efficiency] The Shapley values of all players sum up to the payoff of the grand coalition. That is, $\sum_{j=1}^{n} \shapleyValue_j(\setFunction) = \setFunction(N)$ for any set function \setFunction.
		\item[Symmetry] If two players contribute the same amount to every coalition of other players, then their Shapley values are equal. Formally, for two players $i$ and $j$, if $\setFunction(T \cup \{i\}) = \setFunction(T \cup \{j\})$ for all coalitions $T \subseteq N \setminus \{i, j\}$, then we have $\shapleyValue_i(\setFunction) = \shapleyValue_j(\setFunction)$.
		\item[Null Player] A player that does not contribute to any coalition will get a zero Shapley value. Formally, for a player $j$, if $\setFunction(T \cup \{j\}) = \setFunction(T)$ for all coalitions $T \subset N\setminus\{j\}$, then $\shapleyValue_j(\setFunction) = 0$.
		\item[Additivity] The Shapley value computed from the sum of two set functions is the same as the sum of the Shapley values computed using individual set functions. That is, for any two set functions $\setFunction_1$ and $\setFunction_2$, we have $\shapleyValue_j(\setFunction_1 + \setFunction_2) = \shapleyValue_j(\setFunction_1) + \shapleyValue_j(\setFunction_2)$.
	\end{description}
	Computing the Shapley value for a player has the worst-case time-complexity of $\mathcal{O}(2^n)$. However, efficient approximations exist~\citep{lundberg:2017:shapley-least-squares}.
	
	\subsection{Attributing Marginal Distribution Change}
	
	By slightly abusing the notation, we use the index set $\{1, \dotsc, n\} \eqqcolon N$ to refer to the corresponding variables $\rvX_1, \dotsc, \rvX_n$. Then any coalition $T \subset N$ represents the change set for mechanism changes to the ``old'' causal model. Let $\dist^{\coalition}_{\rvX_k}$ denote the marginal distribution of $\rvX_k$ obtained by marginalising the joint distribution $\dist^{\coalition}_\mrvX$ in the new causal model $\causalModel_T \coloneqq \langle \causalGraph, \dist^{\coalition}_\mrvX \rangle$. Naturally, for $\coalition=\emptyset$, the causal model does not change, i.e. $\causalModel_{\coalition} = \causalModel$ for $\coalition=\emptyset$.
	
	First consider a scenario where the change in the marginal distribution of the target $\rvX_k$ is quantified by the KL divergence. The quantity that we want to attribute to each node is then the KL divergence $\kld{\newdist_{\rvX_k}}{\dist_{\rvX_k}}$. To this end, we define the marginal contribution of a node $\rvX_j$ given a change set \coalition as
	\begin{align*}
		\marginalContribution{\kldSymbol}{j \mid \coalition} &\coloneqq \kld{\dist^{\coalition \cup \{j\}}_{\rvX_k}}{\dist_{\rvX_k}} - \kld{\dist^{\coalition}_{\rvX_k}}{\dist_{\rvX_k}} \;.
	\end{align*}
	In other words, the marginal contribution quantifies how much replacing the causal mechanism of variable $\rvX_j$ contributes to the KL divergence $\kld{\newdist_{\rvX_k}}{\dist_{\rvX_k}}$, given that we have already replaced the causal mechanisms of variables in the change set \coalition. Then the Shapley value of the variable $\rvX_j$ is
	\begin{align}
		\shapleyValue_j(\kldSymbol) \coloneqq \sum_{\coalition \subset N\setminus 
			\{j\}} \frac{1}{n \binom{n-1}{|T|}} \marginalContribution{\kldSymbol}{j \mid \coalition}, \label{eq:shapley-value}
	\end{align}
	which gives the ``fair'' contribution of $\rvX_j$ to the change in the marginal distribution of the target $\rvX_k$ measured in terms of $\kld{\newdist_{\rvX_k}}{\dist_{\rvX_k}}$. Note that the Shapley value contribution $\shapleyValue_j(\kldSymbol)$ can be negative. This is completely reasonable as the marginal contribution $\marginalContribution{\kldSymbol}{j \mid \coalition}$ can be negative; replacing the causal mechanism of a variable can bring the ``new'' marginal distribution of the target closer to the ``old'' marginal distribution $\dist_{\rvX_k}$, thereby lowering the KL divergence relative to not replacing it.
	
	Due to the efficiency property, the Shapley values of all variables will sum up to the KL divergence of the marginal distribution from $\dist_{\rvX_k}$ to $\newdist_{\rvX_k}$, i.e. 
	\begin{align*}
		\sum_{j=1}^{n} \shapleyValue_j(\kldSymbol) = \kld{\newdist_{\rvX_k}}{\dist_{\rvX_k}}\;.
	\end{align*}
	Note that the equality above may not hold strictly when we approximate the Shapley values for players. 
	For the worst case analysis on the approximation error of Shapley values, we refer the interested reader to \citet{charnes:1988:shapley-approx,fatima:2008:linear-shapley-approx}.
	
	Instead of the overall change in the marginal distribution as measured by the KL divergence, we might be interested in the change in some of its property or summary (e.g. mean, median, variance, skew). For instance, it is not uncommon for a retailer to ask, "Why did the mean inventory level go down?", as maintaining an inventory requires upfront investment. Amongst many, that could be due to the change in the distribution of the demand forecast, or some changes in the algorithms (mathematical functions). The definition below provides a rather general way to attribute marginal distribution change.
	
	
	\begin{definition}[Marginal Distribution Change Attribution]
		Let \distSummary denote any functional defined on the marginal distribution. Given a change set \coalition , the marginal contribution of a node $\rvX_j$ to the change in the functional of the marginal distribution of the target $\rvX_k$, i.e. $\Delta \distSummary \coloneqq \distSummary(\newdist_{\rvX_k}) - \distSummary(\dist_{\rvX_k})$, is
		\begin{align*}
			\marginalContribution{\distSummary}{j \mid \coalition} 
			&\coloneqq \distSummary(\dist^{\coalition \cup \{j\}}_{\rvX_k}) - \distSummary(\dist^{\coalition}_{\rvX_k}),
		\end{align*}
		and its Shapley value contribution to $\Delta\distSummary$ is given by
		\begin{align*}
			\shapleyValue_j(\distSummary) \coloneqq \sum_{\coalition \subset N\setminus 
				\{j\}} \frac{1}{n \binom{n-1}{|T|}} \marginalContribution{\distSummary}{j \mid \coalition}.
		\end{align*}
	\end{definition}
	The marginal contribution $\marginalContribution{\distSummary}{j \mid \coalition}$ quantifies how much replacing the causal mechanism of $\rvX_j$ contributes to the change in the summary of $\dist_{\rvX_k}$, given that we have already replaced the causal mechanisms of variables in the change set \coalition. As the marginal contribution of a variable can be negative, the Shapley value contribution $\shapleyValue_j(\distSummary)$ can also be negative. In the example below, we show how to attribute the change in the mean of the target variable to each node.
	
	\begin{example}
		Let $\expectationWithDist{\rvX_k \sim \dist_{\rvX_k}}{\rvX_k}$ denote the mean of the variable $\rvX_k$ under the distribution $\dist_{\rvX}$. The Shapley value contribution of a node $\rvX_j$ to the change in the mean (due to the change in the marginal distribution) of the target node $\rvX_k$ is
		\begin{align*}
			\shapleyValue_j(\expectationSymbol) \coloneqq \sum_{\coalition \subset N\setminus 
				\{j\}} \frac{1}{n \binom{n-1}{|T|}} \marginalContribution{\expectationSymbol}{j \mid \coalition},
		\end{align*}
		where the marginal contribution $\marginalContribution{\expectationSymbol}{j \mid \coalition}$ of $\rvX_j$ given a change set \coalition is defined as
		\begin{align*}
			\marginalContribution{\expectationSymbol}{j \mid \coalition} \coloneqq \expectationWithDist{\rvX_k \sim \dist^{\coalition \cup \{j\}}_{\rvX_k}}{\rvX_k} - \expectationWithDist{\rvX_k \sim \dist^{\coalition}_{\rvX_k}}{\rvX_k}. 
		\end{align*}
		The Shapley values of all nodes sum up to the change in the mean of the target node. That is, the following holds:
		\begin{align*}
			\sum_{j=1}^{n} \shapleyValue_j(\expectationSymbol) = \expectationWithDist{\rvX_k \sim \newdist_{\rvX_k}}{\rvX_k} - \expectationWithDist{\rvX_k \sim \dist_{\rvX_k}}{\rvX_k}.
		\end{align*}
	\end{example}
	Next we discuss practical aspects of distribution change attribution solution we have presented thus far.
	
	\section{Detecting mechanism changes}\label{section:detecting-mechanism-changes}
	To apply our attribution methods, we require a causal graph and its causal conditionals. 
	Existing techniques on sound and complete causal structure learning~\citep{spirtes:00:book, pearl:09:book}, however, can only recover the Markov equivalence class of DAGs assuming faithfulness. With additional assumptions on the data-generating process, it is possible to recover the exact causal graph~\citep{janzing:2017:causality-book,buhlman:2013:l0-penalised-dag-learning}. A promising approach is to combine domain knowledge with conditional independence tests~\citep{mastakouri:2019:causal-discovery}. In some cases, we can also perform controlled randomised experiments to identify the exact DAG from the Markov equivalence class~\citep{eberhardt:2020:causal-discovery}.
	
	Once we have the causal graph, we can then estimate the causal conditionals directly from data using techniques from high-dimensional statistics~\citep{buhlmann:2011:high-dimensional-statistics-book,wainwright:2019:highdim}. Note that we are given two samples of the same size or different sizes (e.g. data from the week before Christmas, and data from the Christmas week). Then for each node $\rvX_j$, we estimate $\dist_{\rvX_j \mid \PA_j}$ from the first sample, and $\newdist_{\rvX_j \mid \PA_j}$ from the second sample. 
	In the context of distribution-change attribution, however, sampling variability can lead to spurious results when we directly plug in the estimated causal conditionals. Even if two samples are drawn from the same joint distribution $\dist_{\rvX_j, \PA_j}$, we will most likely estimate two different causal conditionals (even with regularisation), because of sampling variability. Contrary to the expectation, $\rvX_j$ will then be attributed a non-zero value. 
	
	If we knew that the causal mechanism $\dist_{\rvX_j \mid \PA_j}$ did not change, it would then make sense to learn the causal mechanism from the combined sample, as then the quality of the learned causal conditional will also improve due to the increased sample size. Moreover, we can also directly attribute a zero contribution to the node. This raises the question, "how do we detect causal mechanism changes from two samples?" In other words, are there any statistically testable implications of causal mechanism changes?
	
	There exists a large body of work, e.g. \citet{chakravarti:1967:ks-test,scholz:1987:k-sample-test,snedecor:1989:two-sample-t-test,gretton:2012:kernel-two-sample-test}, on statistical hypothesis test to determine whether the difference between the two distributions is statistically significant from their two samples, each drawn from a separate distribution. Those are not applicable in our setting as we are interested in the difference between the \emph{conditional} distributions, not the marginals. In a recent work, \citet{zhang:2020:causal-discovery-changing-dist} extend the PC algorithm~\cite{spirtes:00:book} for causal discovery from non-stationary or heterogeneous data, where one of the steps involve detecting changing causal mechanisms. Here we adapt their method.
	
	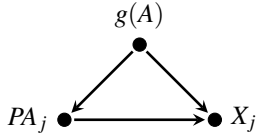
\begin{figure}[tb]
		\centering
		\tikzset{
			var/.style={circle, fill=black, draw=black, inner sep=0pt, minimum size=5pt, node distance=1cm, font=\footnotesize},
			phantom/.style={var, draw=white, fill=white},
			parent/.style={var, draw=red, fill=red},
			descendant/.style={var, draw=blue, fill=blue},
			nondescendant/.style={var, draw=gray, fill=gray},
			directed_edge/.style={->, >=stealth, shorten >=1pt, shorten <=1pt, line width=1pt},
			undirected_edge/.style={shorten >=1pt, shorten <=1pt, line width=1pt}
		}
		\ifincludepdf
			\includegraphics{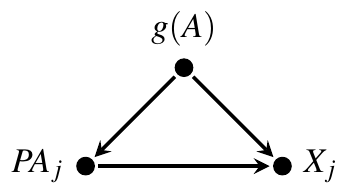}
		\else
			\begin{tikzpicture}
				\node[var, label=left:$\PA_j$] (pa) {};
				\node[phantom, right of=pa] (p) {};
				\node[var, right of=p, label=right:$\rvX_j$] (x) {};
				\node[var, above of=p, label=above:$g(\timeIndex)$] (a) {};
				
				\draw[directed_edge] (a) -- (pa);
				\draw[directed_edge] (a) -- (x);
				\draw[directed_edge] (pa) -- (x);
			\end{tikzpicture}
		\fi
		\caption{The assumed causal graph to detect mechanism changes for a parent-child relationship.}
		\label{fig:detect-mechanism-change}
	\end{figure}
	Assume that causal mechanisms $\dist_{\rvX_j \mid \PA_j}$ can be written as functions of a time or domain index \timeIndex (see Figure~\ref{fig:detect-mechanism-change}). If the causal graph is induced by a functional causal model, then the quantities such as functional models, noise levels, etc that may change over time or across domains can be written as functions of \timeIndex. Under these assumptions, if the causal mechanism $\dist_{\rvX_j \mid \PA_j}$ remains the same across various values of \timeIndex, then the conditional independence test $\rvX_j \indep \timeIndex \mid \PA_j$ suffices to detect changes to the causal mechanism $\dist_{\rvX_j \mid \PA_j}$.
	
	Let $D_t$ denote the $\by{m_t}{n}$ matrix containing sample from time (or domain) $t \in \{1, 2\}$, and $D$ denote the $\by{m}{n}$ matrix obtained by vertically concatenating $D_t$s, where $m=m_1+m_2$. 
	That is, we have
	\begin{align*}
		D \coloneqq 
		\begin{bmatrix}
			D_1\\
			D_2
		\end{bmatrix}
	\end{align*}
	We can construct \timeIndex directly from data. The key idea is to assign the same value of \timeIndex to the units in the sample from the same time (or domain) $t$. 
	For clarity, with a slight abuse of notation, we interchangeably use \timeIndex for a variable as well as the data vector. Each entry $a_i$ of \by{m}{1} vector \timeIndex is assigned
	\begin{align*}
		a^{(i)} \coloneqq 
		\begin{cases}
			+1, &\text{if } i \leq m_1,\\
			-1, &\text{otherwise}.
		\end{cases}
	\end{align*}
	Using the columns from the combined data matrix $D$ and index vector \timeIndex, we then test if each variable $\rvX_j$ is conditionally independent of \timeIndex given its direct parents $\PA_j$ in the causal graph, i.e. $\rvX_j \indep \timeIndex \mid \PA_j$. If the conditional independence holds, then the causal mechanism $\dist_{\rvX_j \mid \PA_j}$ did not change across various values of \timeIndex. On the other hand, if $\rvX_j$ is dependent on \timeIndex given $\PA_j$ then its causal mechanism $\dist_{\rvX_j \mid \PA_j}$ also changes with \timeIndex. 
	
	Note that the functional relationships between variables $\rvX_j$ and time (or domain) index \timeIndex is unknown. It is therefore important to use a non-parametric conditional independence test. In this work, we use kernel-based conditional independence test~\citep{zhang:2011:kci-test}.
	
	\section{Related Work}\label{section:related}
	Path-specific effect of \rvX and \rvY via path $\pi$ is the degree to which an interventional-change in \rvX would change the marginal distribution of \rvY if that change were to be transmitted only via $\pi$. If an indicator (or context) variable $A$ represents samples from distributions $P$ and $\tilde{P}$, then computing the path-specific effect of $A$ on any node via the direct path simply measures the distance between the ``old'' causal mechanism and the ``new'' causal mechanism of the node. Arguably, we then capture the causal influence of external factors (abstracted by $A$) to the node~\citep{janzing:2013:direct-arrow-strength}. To discuss the relation to the strength of causal arrows defined in \citet{janzing:2013:direct-arrow-strength}, we need to label the two distributions by an additional variable $V$ attaining values $v$ and $\bar{v}$ with edges to all $\rvX_j$ whose mechanisms change. Their strengths would be $\kld{\dist_{\rvX_j|\PA_j}} {[p(v) \dist_{\rvX_j \mid \PA_j} + p(\bar{v}) \dist_{\rvX_j|\PA_j} ]}$. 
	
	Most feature-based interpretability techniques assume that features independently co-cause the target~\citep{lundberg:2017:shapley-least-squares,janzing:2020:feature-attribution}. In particular, they consider how the marginal distribution of the target changes w.r.t. to interventional changes in the features. They cannot attribute causal mechanism changes, as one must assume that the causal mechanism of the target does not change to make them work for multiple datasets.
	
	\citet{kulinski:2020:feature-shift-detection} perform a statistical test for the distance between conditional distribution of each feature given other features from two samples to assign blame of distribution shift to a subset of features. As direct causes are not considered in conditioning, they do not answer the "why?" question.
	
	In causal discovery from multiple contexts~\citep{mooij:2020:JCI}, they find the union of causal graphs in each dataset (or context) by jointly modelling the context variables ($A_1, \dotsc, A_k$) and observed variables ($\rvX_1, \dotsc, \rvX_n$). While they also work on samples from different distributions and hence might appear related, these are two different problems.
	
	Mechanism changes can also be represented with stochastic interventions~\citep{pearl:09:book, correa:2020:stochastic-interventional-calculus}. We briefly discussed the connection in the last paragraph of Section~\ref{section:mechanism-changes}. Computing the effect of a stochastic intervention, however, does not tell us how much the corresponding mechanism change contributed to a target quantity summarising distribution change.
	
	Overall, existing methods are not suitable for distribution change attribution, where our goal is to attribute  a target quantity summarising distribution change (e.g. change in mean, KL divergence) to change in causal mechanism (conditional distribution of a node given its direct causes) of each variable.
	
	\section{Experiments}\label{section:experiments}
	In experiments, we study the performance of our approach for attributing the change in the marginal distribution, and its application to real-world data. In particular, through simulations, we study the performance of our attribution method when we learn causal mechanisms from samples w.r.t. sample size and magnitude of causal mechanism change. On real-world data, we asses whether results are sensible.
	
	\subsection{Simulations}
	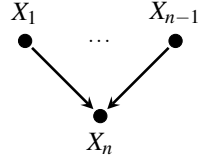
\begin{figure}[tb]
		\centering
		\tikzset{
			var/.style={circle, fill=black, draw=black, inner sep=0pt, minimum size=5pt, node distance=1cm, font=\footnotesize},
			phantom/.style={var, draw=white, fill=white},
			parent/.style={var, draw=red, fill=red},
			descendant/.style={var, draw=blue, fill=blue},
			nondescendant/.style={var, draw=gray, fill=gray},
			directed_edge/.style={->, >=stealth, shorten >=1pt, shorten <=1pt, line width=1pt},
			undirected_edge/.style={shorten >=1pt, shorten <=1pt, line width=1pt}
		}
		\ifincludepdf
			\includegraphics{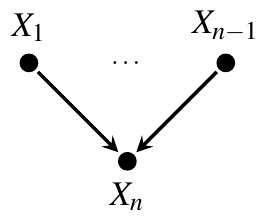}
		\else
		\begin{tikzpicture}
			\node[var, label=above:$\rvX_1$] (X1) {};
			\node[phantom, right of=X1] (Xcontd) {$\dotso$};
			\node[var, right of=Xcontd, label=above:$\rvX_{n-1}$] (Xlast) {};
			\node[var, below of=Xcontd, label=below:$\rvX_n$] (Xn) {};
			
			\draw[directed_edge] (X1) -- (Xn);
			\draw[directed_edge] (Xlast) -- (Xn);
		\end{tikzpicture}
		\fi
		\caption{The causal graph used in simulations, where independent inputs $\rvX_1, \dotsc, \rvX_{n-1}$ co-cause the target $\rvX_n$.}
		\label{fig:simulation-causal-graph}
	\end{figure}
	
	We consider a setting where $n-1$ independent input variables $\rvX_1, \dotsc, \rvX_{n-1}$ co-cause a target variable $\rvX_n$. Their underlying causal graph is shown in Figure~\ref{fig:simulation-causal-graph}. We choose this simple, yet representative, causal graph for simulation because computing the Shapley values analytically for summaries of distributional changes is not trivial for rather complex graphs.
	
	Let $\rvN_{w} \sim \mathcal{N}(\mu_w, 1)$ denote an independent Gaussian noise with mean $\mu_w$ and unit variance for each $w \in \{1, \dotsc, n\}$. Suppose that their causal graph is induced by the following structural assignments: 
	\begin{align*}
		\rvX_w &\coloneqq \rvN_w \text{ for } w \in \{1, \dotsc, n-1\} \text{ and } \\
		\rvX_n &\coloneqq \rvX_1 + \dotso + \rvX_{n-1} + \rvN_n.
	\end{align*}
	We refer to the structural causal model (SCM) above by \scm. 
	Suppose that their joint distribution $\Pr_{\rvX_1, \dotsc, \rvX_n}$ changes to $\tilde{\Pr}_{\rvX_1, \dotsc, \rvX_n}$ due to the changes in the structural assignments. In the new SCM $\tilde{\scm}$, we have the following assignments:
	\begin{align*}
		\rvX_w &\coloneqq \rvN_w + \lambda_w \text{ for } w \in \{1, \dotsc, n-1\} \text{ and } \\
		\rvX_n &\coloneqq \rvX_1 + \dotso + \rvX_{n-1} + \rvN_n + \lambda_n,
	\end{align*}
	where $\lambda_w$ is a scalar that shifts the mean of the corresponding variable $\rvX_w$ for each $w \in \{1, \dotsc, n\}$, and further defined as
	\begin{align}
		\lambda_w \coloneqq \begin{cases}
			\lambda &\text{ if } S_w=1\\
			0 &\text{ otherwise, }
		\end{cases}
	\end{align}
	where $S_w$ is a Bernoulli random variable with a probability of success $p=0.5$. That is, $S_w$ determines whether the causal mechanism of the corresponding variable $\rvX_w$ is potentially subject to change. If $S_w=1$, then the value of $\lambda$ subsequently dictates the \emph{magnitude} of the change in the causal mechanism of $\rvX_w$. Note that even if $S_w=1$, the causal mechanism of corresponding variable $\rvX_w$ does not change if $\lambda=0$. With rejection sampling, we ensure that at least one causal mechanism changes, i.e. $\lambda_w \neq 0$ for at least one $w \in \{1, \dotsc, n\}$. This way, we can change the causal mechanism of a random subset of variables through $S_w$, and study the performance of our attribution method w.r.t $\lambda$.
	
	Due to changes in the causal mechanisms of variables, the marginal distribution of the target $\rvX_n$ also changes:
	\begin{align*}
		\scm &: \rvX_n \sim \mathcal{N}(\mu_1 + \dotso + \mu_n, n)\\
		\tilde{\scm}&: \rvX_n \sim \mathcal{N}(\mu_1 + \dotso + \mu_n + \lambda_1 + \dotso + \lambda_n, n).
	\end{align*}
	Let $\Pr_{\rvX_n}$ and $\tilde{\Pr}_{\rvX_n}$ denote the marginal distributions of $\rvX_n$ in SCMs \scm and $\tilde{\scm}$ respectively. We measure the change in the marginal distribution of $\rvX_n$ by the difference in its mean, i.e.
	\begin{align*}
		\Delta\expectationSymbol &\coloneqq \expectationWithDist{\rvX_n \sim \newdist_{\rvX_n}}{\rvX_n} - \expectationWithDist{\rvX_n \sim \dist_{\rvX_n}}{\rvX_n}\\
		&= \lambda_1 + \dotso + \lambda_n .
	\end{align*}
	With some algebraic manipulation, we can show that the contribution of each variable $\rvX_w$---due to the change in its causal mechanism---to $\Delta\expectationSymbol$ is then given by
	\begin{align}
		\phi_{w}(\expectationSymbol) = \lambda_w
	\end{align}
	These closed-form expressions provide the ground truth for our evaluation. As it should, we see that the following holds: 
	\[
	\phi_{1}(\expectationSymbol) + \dotso + \phi_{n}(\expectationSymbol) = \Delta \expectationSymbol .
	\]
	
	First we generate two set of samples of same size from the two SCMs \scm and $\tilde{\scm}$ stated above. 
	From each sample, we learn (estimate) the SCM assuming that the causal graph is known. As the SCM has an additive unobserved noise term, it is possible to estimate both the function and the noise from data with regression. We generate a sample from the joint distribution induced by the learned SCM. Then we estimate the mean of the marginal distribution by the sample average. Finally, we compute the Shapley value of each variable $\rvX_w$. To measure the quality of the estimated Shapley values against the ground truth, we use the $\ell_1$ norm, otherwise known as Manhattan distance. 
	
	\begin{figure}[tb]
		\ifincludepdf
			\includegraphics{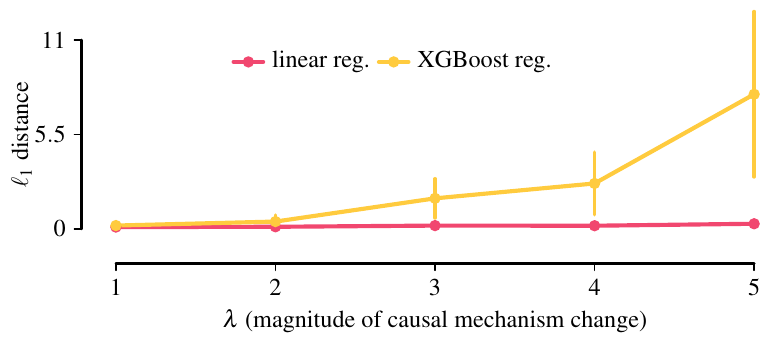}
		\else
			\begin{tikzpicture}
				\begin{axis}[
					clean mark, 
					width=\columnwidth, 
					height=3.5cm, 
					xlabel=$\lambda$ (magnitude of causal mechanism change), 
					ylabel=$\ell_1$ distance,
					ylabel style = {yshift=30pt},
					xlabel style = {yshift=-15pt},
					ymin=0,
					ymax=11,
					ytick={0, 5.5, 11},
					legend columns=2,
					legend style = {
						draw=none,
						at={(0.7,1.0)}
					},
					]
					\addplot+[very thick, color4, mark=*, mark options={fill=color4}, error bars/.cd, error bar style={color4, line width=1pt}, error mark=line, error mark options={mark size=1pt}, y dir=both, y explicit] table[x=strength,y=mean_linear, header=true, col sep=comma, y error=std_linear] {data/simulation-strength.csv};
					\addplot+[very thick, color2, mark=*, mark options={fill=color2}, error bars/.cd, error bar style={color2, line width=1pt}, error mark=line, error mark options={mark size=1pt}, y dir=both, y explicit] table[x=strength,y=mean_xgboost, header=true, col sep=comma, y error=std_xgboost] {data/simulation-strength.csv};
					\legend{linear reg., XGBoost reg.};
				\end{axis}
			\end{tikzpicture}
		\fi
		\caption{The $\ell_1$ distance between the ground truth, and the estimated Shapley values when the underlying linear structural causal model is estimated with the linear regression model versus the graident boosted trees regression model at various values of $\lambda$. The standard error bars for the linear regression model are invisble as they are too narrow.}
		\label{fig:simulation-strength}
	\end{figure}
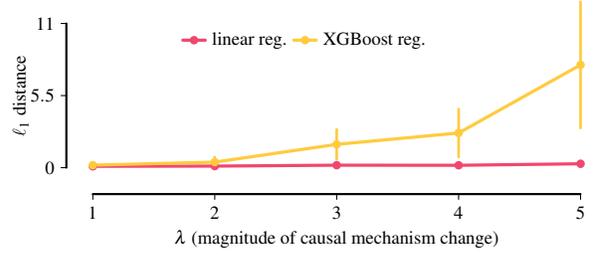
	
	First we study the performance of our attribution method against the magnitude parameter $\lambda$. To this end, for a given value of $\lambda$, we generate 100 pair of SCMs (\scm, $\tilde{\scm}$) with $\mu_w$ chosen according to the Uniform distribution $\mathcal{U}(-5, 5)$ for each $w \in \{1, \dotsc, n\}$, where $n$ is chosen uniformly randomly from $\{2, 3, 4, 5\}$. From each SCM in the pair (\scm, $\tilde{\scm}$), we generate 100 samples, each containing 1000 observations. Note that we learn the SCM from each sample, and then estimate the Shapley values from the sample drawn from the learned SCM, which we repeat 100 times. Therefore we report the average $\ell_1$ distance (with standard error) over $100\times100\times100=1\,000\,000$ pairs of samples in Figure~\ref{fig:simulation-strength} at various values of $\lambda$. With a right regression model (linear regression), the estimated Shapley values are very close to the ground truth regardless of the magnitude of causal mechanism change $\lambda$---indicated by close to zero $\ell_1$ distance. With gradient boosted trees (from \texttt{xgboost} python package with default hyperparameters and 100 trees), however, the estimated Shapley values, on average, differ from the ground truth as $\lambda$ increases. This is expected as inferring the right model is harder if function classes are less restricted a priori.
	Therefore, the Shapley values estimated using a non-linear function will deviate from the ground truth compared to that from a linear function. We also observe that the uncertainty in the Shapley value estimation increases if model estimation is not accurate. This is indicated by wide error bars for the gradient boosted trees compared to narrow error bars, that are invisible in the figure, for the linear regression.
	
\begin{figure}[tb]
		\centering
		\ifincludepdf
			\includegraphics{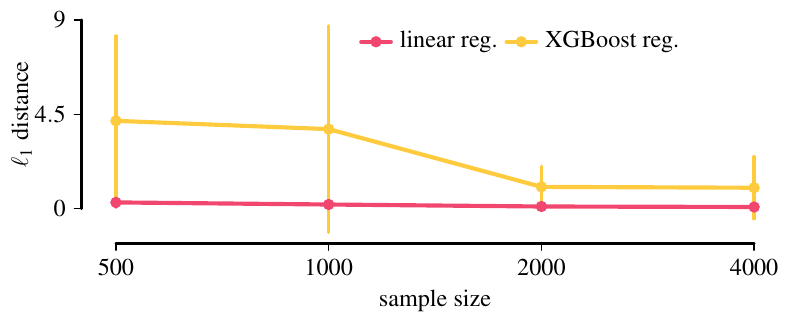}
		\else
			\begin{tikzpicture}
				\begin{axis}[
					clean mark, 
					width=\columnwidth, 
					height=3.5cm, 
					xlabel=sample size, 
					ylabel=$\ell_1$ distance,
					ylabel style = {yshift=30pt},
					xlabel style = {yshift=-15pt},
					ymin=0,
					ymax=9,
					ytick={0, 4.5, 9},
					xtick={500, 1000, 2000, 4000},
					symbolic x coords = {500, 1000, 2000, 4000},
					legend columns=2,
					legend style = {
						draw=none,
						at={(0.9,1.0)}
					},
					]
					\addplot+[very thick, color4, mark=*, mark options={fill=color4}, error bars/.cd, error bar style={color4, line width=1pt}, error mark=line, error mark options={mark size=1pt}, y dir=both, y explicit] table[x=sample_size,y=mean_linear, header=true, col sep=comma, y error=std_linear] {data/simulation-samplesize.csv};
					\addplot+[very thick, color2, mark=*, mark options={fill=color2}, error bars/.cd, error bar style={color2, line width=1pt}, error mark=line, error mark options={mark size=1pt}, y dir=both, y explicit] table[x=sample_size,y=mean_xgboost, header=true, col sep=comma, y error=std_xgboost] {data/simulation-samplesize.csv};
					\legend{linear reg., XGBoost reg.};
				\end{axis}
			\end{tikzpicture}
		\fi
		\caption{The $\ell_1$ distance between the ground truth, and the estimated Shapley values when the underlying linear structural causal model is estimated with linear regression versus XGBoost regression at various sample sizes.}
		\label{fig:simulation-samplesize}
	\end{figure}
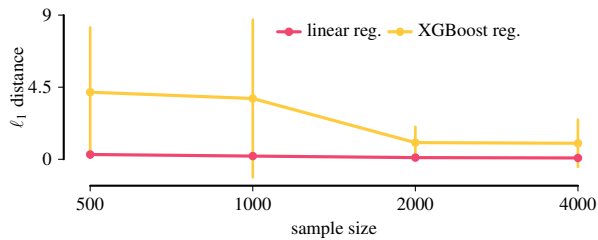

	In Figure~\ref{fig:simulation-samplesize}, we show the result when we vary the sample size, but randomly choose the magnitude parameter $\lambda$ according to $\mathcal{U}(1, 5)$ for each SCM pair. We observe that the Shapley values from linear regression model is close to the ground truth even at a relatively small sample size of 500---with an average $\ell_1$ distance of 0.29 and standard error of 0.21. While the performance of XGBoost regression model certainly improves with increasing sample size, its performance does not match the linear regression model.
	
	\subsection{Case Study}
	Next we present a case study on the Adult Census Income dataset\footnote{\url{http://archive.ics.uci.edu/ml/datasets/Adult}} where we use our proposal to identify the drivers of the difference in the income distribution between men and women. 
	
	The dataset contains 32,561 records from the census on annual income in the United States from 1994. In addition to whether the annual income of an individual is greater than fifty thousand USD, it contains 14 other socio-economic attributes. We consider a subset of non-redundant attributes for analysis, namely education and occupation, that directly affect income as well as act a proxy of income for other attributes. After removing the rows with missing values, we end up with 30,718 rows.
	
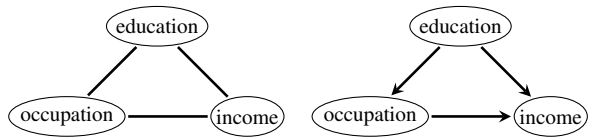
\begin{figure}[tb]
		\centering
		\tikzset{
			var/.style={ellipse, draw=black, inner sep=-2pt, minimum size=15pt, node distance=1.2cm, font=\footnotesize},
			phantom/.style={var, draw=white, fill=white},
			parent/.style={var, draw=red, fill=red},
			descendant/.style={var, draw=blue, fill=blue},
			nondescendant/.style={var, draw=gray, fill=gray},
			directed_edge/.style={->, >=stealth, shorten >=1pt, shorten <=1pt, line width=1pt},
			undirected_edge/.style={shorten >=1pt, shorten <=1pt, line width=1pt}
		}
		\begin{minipage}{0.5\columnwidth}%
			\ifincludepdf
				\includegraphics{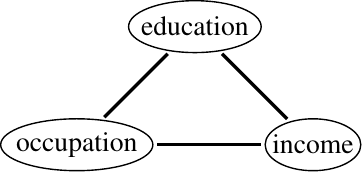}
			\else
				\begin{tikzpicture}
					\node[var] (education) {education};
					\node[phantom, below of=education] (phantom) {};
					\node[var, left of=phantom] (occupation) {occupation};
					\node[var, right of=phantom] (income) {income};
					
					\draw[undirected_edge] (education) -- (occupation);
					\draw[undirected_edge] (education) -- (income);
					\draw[undirected_edge] (occupation) -- (income);
				\end{tikzpicture}
			\fi
		\end{minipage}%
		\begin{minipage}{0.5\columnwidth}%
			\ifincludepdf
				\includegraphics{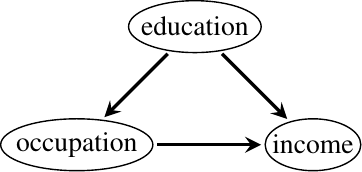}
			\else
				\begin{tikzpicture}
					\node[var] (education) {education};
					\node[phantom, below of=education] (phantom) {};
					\node[var, left of=phantom] (occupation) {occupation};
					\node[var, right of=phantom] (income) {income};
					
					\draw[directed_edge] (education) -- (occupation);
					\draw[directed_edge] (education) -- (income);
					\draw[directed_edge] (occupation) -- (income);
				\end{tikzpicture}
			\fi
		\end{minipage}%
		\caption{(left) Skeleton graph discovered using PC algorithm on a subset of variables from the Adult Census Income dataset. (right) Causal graph derived from the skeleton by orienting the undirected edges using domain knowledge.}
		\label{fig:adult-causal-graph}
	\end{figure}

	As the number of variables is small, we combine causal discovery with domain knowledge to construct the causal graph. First, from the combined records of men and women, we discover the skeleton graph using the PC algorithm~\citep{spirtes:00:book} with kernel-based conditional independence test~\citep{zhang:2011:kci-test} at a significance level of $0.2$ (see Figure~\ref{fig:adult-causal-graph} left). At a higher significance level, the PC algorithm gives us a denser skeleton. This way we minimise the chances of omitting dependencies. 
	We then orient the edges in the skeleton using domain knowledge. Note that changing the occupation will ``mainly'' lead to the change in the annual income, not the other way around.
	Changing the education not only affects the occupation, but also the income~\citep{heckman:2018:education-income}, e.g. passive income through smart investments, side incomes, etc. Therefore, education confounds both occupation and income. We then have a causal graph as shown in Figure~\ref{fig:adult-causal-graph} (right).
	
	Since our goal is to identify the drivers of difference in the income distribution between men and women, as a sanity check, we perform a two-sample test to determine whether the income distribution is, indeed, different between men and women in the dataset. Under the null hypothesis that incomes of men and women come from the same distribution, the Kolmogorov-Smirnov two-sample test yields a $p$-value of $1.38\times 10^{-234}$. Since the $p$-value is extremely low, we can safely reject the null hypothesis.
	
	\begin{figure}[tb]
		\centering
		\ifincludepdf
			\includegraphics{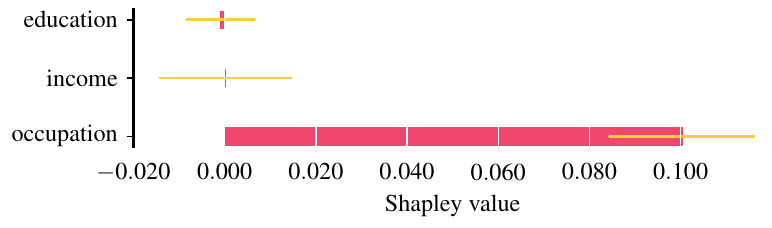}
		\else
			\begin{tikzpicture}
				\begin{axis}[
					clean xbar, 
					width=\columnwidth, 
					enlarge y limits=0.1, 
					bar width=0.5em,
					height=3cm, 
					xtick={-0.02, 0.0, ..., 0.12},
					xmin=-0.02,
					xmax=0.12,
					ytick=data, 
					xlabel=Shapley value,
					x label style={yshift=-1.2em},
					symbolic y coords={occupation, income, education, cap-gain},
					x axis line style={draw=none},
					legend style={at={(1.0,1.0)}, draw=none},
					legend image post style={scale=0.4},
					x tick label style={
						/pgf/number format/.cd,
						fixed,
						fixed zerofill,
						precision=3,
						/tikz/.cd
					},
					scaled x ticks=false,
					]
					\addplot+[error bars/.cd, error bar style={color2, thick,}, error mark options={mark size=0pt,line width=0pt}, x dir=both, x explicit] table[y=node, x=mean-contrib, col sep=comma, header=true, x error expr=\thisrow{conf-max}-\thisrow{mean-contrib}] {data/adult_result.csv};
				\end{axis}
			\end{tikzpicture}
		\fi
		\caption{Shapley value contribution of each variable (due to the potential difference in its causal mechanism) to the difference in the mean annual income between men and women ($\mu_{\textrm{men}}-\mu_{\textrm{women}}$). Each horizontal bar represents the mean Shapley value contribution, and each line represents the bias-corrected and accelerated bootstrap confidence interval at a 95\% confidence level, over 100 resamples. The horizontal bars of education and income are almost invisible as their mean Shapley value contributions are close to zero.}
		\label{fig:adult-result}
	\end{figure}
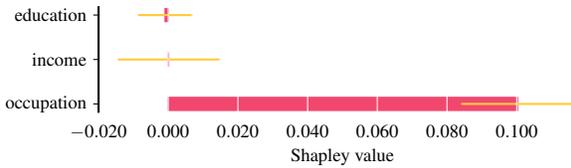
	
	All three variables are categorical. In particular, the target variable (income) is binary (>50K?). We use the empirical distribution as the causal mechanism of the root node (education).
	For a non-root node (occupation and income), we learn its causal mechanism using a XGBoost classifier with 100 gradient boosted trees. 
	To detect causal mechanism changes, we use kernel-based conditional independence test~\citep{gretton:2012:kernel-two-sample-test} with delta kernel at a significance level of 0.05. We would like to attribute the difference in the \emph{mean} annual income between men and women.

\begin{figure}
		\centering
		\ifincludepdf
			\includegraphics{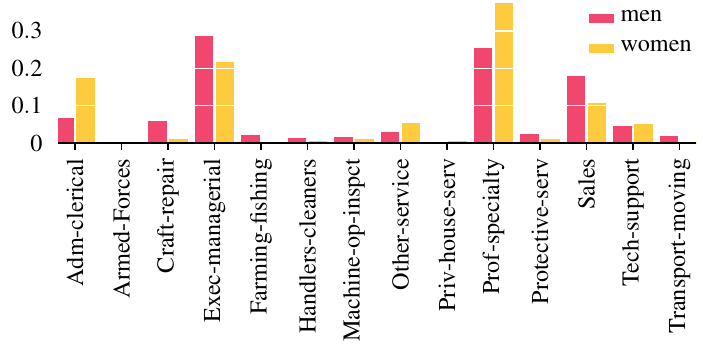}
		\else
			\begin{tikzpicture}
				\begin{axis}[
					clean ybar, 
					ymin=0.0,
					width=\columnwidth, 
					enlarge x limits=0.028, 
					bar width=0.5em,
					height=3cm, 
					xtick=data, 
					symbolic x coords={Adm-clerical,Armed-Forces,Craft-repair,Exec-managerial,Farming-fishing,Handlers-cleaners,Machine-op-inspct,Other-service,Priv-house-serv,Prof-specialty,Protective-serv,Sales,Tech-support,Transport-moving},
					x tick label style={rotate=90},
					legend style={at={(1.01,1.03)}, draw=none},
					legend image post style={scale=0.4},
					]
					\addplot table[x=occupation, y=male, col sep=comma, header=true] {data/cpt-occupation.csv};
					\addplot table[x=occupation, y=female, col sep=comma, header=true] {data/cpt-occupation.csv};
					\legend{men, women};
				\end{axis}
			\end{tikzpicture}
		\fi
		\caption{Conditional probability distribution of occupation given ``Bachelor'' educated men versus women.}
		\label{fig:adult-masters}
	\end{figure}
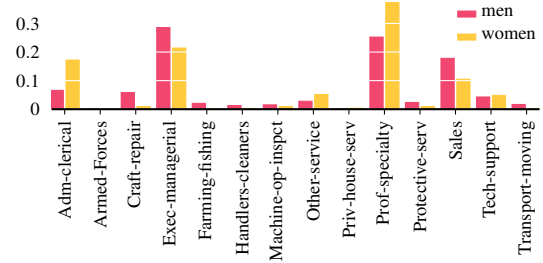

	The result of our method is shown in Figure~\ref{fig:adult-result}. We find that the change in the causal mechanism of occupation, $\dist_{\text{occupation} \mid \text{education}}$, is the main driver for the difference in the mean annual income between men and women. This is also often cited in public discourse as a reason why we need more women participation in labour market to empower them economically~\citep{giuliano:2017:gender-history}.\!\footnote{A comprehensive review of literature on this topic along with data and visualisation is available at \url{https://ourworldindata.org/female-labor-supply}.} Educational choices of men and women differ---science-related subjects are attended mostly by men than women~\citep{wang:2017:gender-gap-stem,tellhed:2016:gender-gap-stem}. In UC Berkeley gender bias study from 1975, for instance, it was found that, compared to men, women tended to apply to departments that are more crowded, less well funded, and that frequently offer poorer professional employment prospects~\citep{bickel:1975:berkeley-admission-bias}. 
	Graduates of science-related subjects are also known to earn more than the others~\citep{deming:2018:stem-career}.
	Moreover, women often take non-professional responsibilities, such as parenting and caring family relatives that affect their career and earnings~\citep{jolly:2014:gender-career,budig:2006:gender-income}. Therefore, income distribution of men and women differ even when they have the same level of education. As a representative case, we show the empirical conditional probability distribution of occupation given ``Bachelor'' educated men versus women in Figure~\ref{fig:adult-masters}, which also corroborates our attribution result. The finding that the subject of education plays a significant role on occupation and income would deserve further studies with more detailed data sets which goes beyond the scope of the present paper.
	
	\section{Discussion and Conclusions}\label{section:conclusions}
	We presented a formal approach to identify the drivers of distribution change using graphical causal models. The key idea is that, given a causal graph, we can factorise the joint distribution into independent causal conditionals. Any change in the joint distribution, or marginal distribution of any target variable thereof, can then be attributed to changes in some of the causal conditionals. We illustrated our method on both simulated and real-world datasets. 
	
	In Section~\ref{section:detecting-mechanism-changes}, we showed how to detect causal mechanism changes from data given the underlying causal graph using conditional independence tests. In many tasks---e.g. exploratory data analysis, designing and evaluating robust models---knowing those conditionals that change is already sufficient. One might then ask, "Why do we need to quantify the contribution from each mechanism?" Causal mechanisms always change in a system of a large number of variables that are embedded in a changing environment. Supply chain is one such example where continuous deployments of new changes in constituent subsystems (e.g. forecasting, buying) are common. It is too costly (e.g. time, personnel) to look at all variables whose mechanisms change. Quantifying how much each variable contributed to the change allows us to focus on a few most relevant variables.
	
	Finally, our attribution approach requires a causal graph, which may not be identifiable from observational data. If the causal graph is not identifiable, the Shapley values will not be identifiable as well. Therefore, this question on the robustness of our attribution approach to causal graph misspecification deserves further research.
	
	\section*{Acknowledgements}
	The authors thank Dr. David Afshartous for useful comments.
	
	\balance
	\bibliographystyle{abbrvnat}
	\bibliography{paper}
\end{document}